\documentclass[3p,times,procedia]{elsarticle}
\usepackage{nupha_ecrc}

\usepackage{xcolor}
\definecolor{lcolor}{rgb}{0.5,0,0}
\definecolor{citcolor}{rgb}{0,0.3,0.0}

\usepackage[breaklinks,colorlinks,urlcolor=blue,citecolor=citcolor,linkcolor=lcolor]{hyperref}

\volume{00}

\firstpage{1}

\journalname{Nuclear Physics A}

\runauth{T. Lappi}

\jid{nupha}

\jnltitlelogo{Nuclear Physics A}

\usepackage{amssymb}
\usepackage{amsmath}

\usepackage[figuresright]{rotating}

\newcommand{\rt}{{\mathbf{r}_T}}

\newcommand{\xt}{{\mathbf{x}_T}}
\newcommand{\bt}{{\mathbf{b}_T}}

\newcommand{\yt}{{\mathbf{y}_T}}

\newcommand{\pt}{{\mathbf{p}_T}}

\newcommand{\ptt}{p_T} 

\newcommand{\nabt}{\boldsymbol{\nabla}_T}

\newcommand{\ud}{\, \mathrm{d}}

\newcommand{\tr}{\, \mathrm{Tr} \, }

\newcommand{\nc}{{N_\mathrm{c}}}

\newcommand{\nr}[1]{(\ref{#1})}

\newcommand{\qs}{Q_\mathrm{s}}

\newcommand{\fig}{Fig.~}

\newcommand{\eq}{Eq.~}

\begin{document}

\begin{frontmatter}



\dochead{}

\title{Initial state azimuthal anisotropies in small collision systems}

\author[label1,label2]{T. Lappi}
\address[label1]{Department of Physics,  P.O. Box 35, 40014 University of Jyv\"askyl\"a, Finland }
\address[label2]{Helsinki Institute of Physics, P.O. Box 64, 00014 University of Helsinki, Finland}


\begin{abstract}
Strong multiparticle azimuthal correlations have recently been observed in high energy proton-nucleus collisions. 
While final state collective effects can be responsible for many of the observations,
the domain structure in the classical color field of a high energy nucleus also naturally leads to such correlations. We describe  recent calculations of the momentum space 2-particle cumulant azimuthal anisotropy coefficients $v_n\{2\}$, $n=2,3,4$ from fundamental representation Wilson line distributions describing the high energy nucleus. We find significant differences between Wilson lines from the MV model and from JIMWLK evolution. We also discuss the relation of this calculation to earlier work on the ridge correlation obtained in the ``glasma graph'' approximation, and to the ``color electric field domain model.''
\end{abstract}


\end{frontmatter}



\section{Introduction}

Smaller reference systems are important for  understanding  quark gluon plasma production in high energy nucleus-nucleus collisions. Proton-proton and proton nucleus collisions probe the limits of applicability of hydrodynamics and more directly reveal nontrivial initial state collective effects. Here  we will discuss  the extreme limit of a very dilute projectile colliding on a target of nonperturbatively dense QCD color fields; our discussion follows  that of Refs.~\cite{Lappi:2015vha,Lappi:2015vta}.

Azimuthal anisotropies  in high energy nuclear collisions are usually parametrized by harmonic coefficients of the spectra of produced particles, commonly denoted as $v_n$.   For the purposes of this discussion, these are always determined from multiparticle correlations which are usually, depending on the exact analysis method, long range in rapidity. 
A very simple causality argument tells us that these  correlations originate in the earliest stage of the collision process.  The transverse geometry is naturally a very long range correlation; particle production at all rapidities is sensitive to the positions of the nucleons inside the colliding nuclei. The effect of collective flow is to transform this initially position space correlation into momentum space through the forces caused by pressure anisotropies. What we discuss here are another source of azimuthal anisotropies that can be generated directly in momentum space in the particle production process.

\section{Correlations from color field domains in the target}

The calculation of particle production in the forward dilute-dense limit in the CGC formalism corresponds to a very simple physical picture. Collinear partons from the dilute probe (which can be described in terms of usual parton distribution functions) pass through the color field of the target. They obtain a transverse momentum kick from the target color electric field and emerge as real produced particles. The color field of the target consists of domains of size $1/\qs$, which corresponds to the  typical intrinsic transverse momentum of the target gluons. If two uncorrelated particles from the probe are in the same color state (which happens with a probability $\sim 1/\nc^2$) and pass though the same domain (the probability for this being $\sim 1/\left(S_\perp \qs^2\right)$ where $S_\perp$ is the area of the probe), the transverse momentum kick is in the same direction and they become correlated. The parametric estimate for the correlation strength $\sim 1/\left(\nc^2 \qs^2 S_\perp\right)$ means that, contrary to the hydrodynamical explanation, these correlations are relatively stronger in small systems. The same structure of color field domains is present also in the dense-dense collision system and leads to parametrically similar correlations~\cite{Gelis:2009wh}, although the calculation becomes somewhat more complicated.

In Ref.~\cite{Lappi:2015vha} the azimuthal correlations in this picture in a homogenous (in the statistical sense) infinite target were calculated in the following simple setup. The passage of the probe particle through the target can be described, as usually in the CGC formalism, in terms of an  eikonal Wilson line in the color field
\begin{equation}
V(\xt) = P \exp\left\{i g \int \ud x^- A^+_\mathrm{cov}(\xt,x^-) \right\} ,
\end{equation}
where $P$ denotes path ordering along the lightlike trajectory of the probe. As is clear from the above discussion, it is essential for the correlation to localize the particles in the probe. In Ref.~\cite{Lappi:2015vha} this is done by explicitly imposing a Gaussian weight function (corresponding to the Wigner distribution of a minimal Gaussian wave packet) independently for each of the particle in the projectile.  The single inclusive particle spectrum for a fixed configuration of target color fields then becomes:
\begin{equation}
\frac{\ud N}{\ud^2\pt}\propto 
\int\limits_{\xt,\yt}e^{- i \pt \cdot(\xt-\yt)} 
e^{\frac{-(\xt-\bt)^2}{2B}}
e^{\frac{-(\yt-\bt)^2}{2B} }
\frac{1}{\nc} \tr V_\xt^\dag V_\yt.
\end{equation}\label{eq:sinc}
The two particle correlation can be calculated by squaring the single particle distribution \nr{eq:sinc} and then averaging over the color field configurations of the target. It thus depends on a 4-point function of the Wilson lines $V$.
It is then decomposed in a Fourier-series in the azimuthal angle to obtain the two particle cumulant anisotropy  coefficients $v_n\{2\}$, following the procedure explained e.g. in Ref.~\cite{Chatrchyan:2013nka}. 

To evaluate the target color field expectation values one needs a probability distribution for the Wilson lines $V(\xt)$; this distribution contains the physics of the gluon correlations in the target color fields. In Ref.~\cite{Lappi:2015vha} the Wilson lines are obtained from two different distributions. The first one is the McLerran-Venugopalan (MV) model~\cite{McLerran:1994ni} of independent color charges with a Gaussian distribution. The second one is the result from  leading order running coupling  JIMWLK nonlinear renormalization group evolution (see \cite{Lappi:2012vw}) describing the dependence of the probability distribution on $x$, with the MV model as an initial condition.

\begin{figure}
\centerline{
\includegraphics[width=0.33\textwidth]{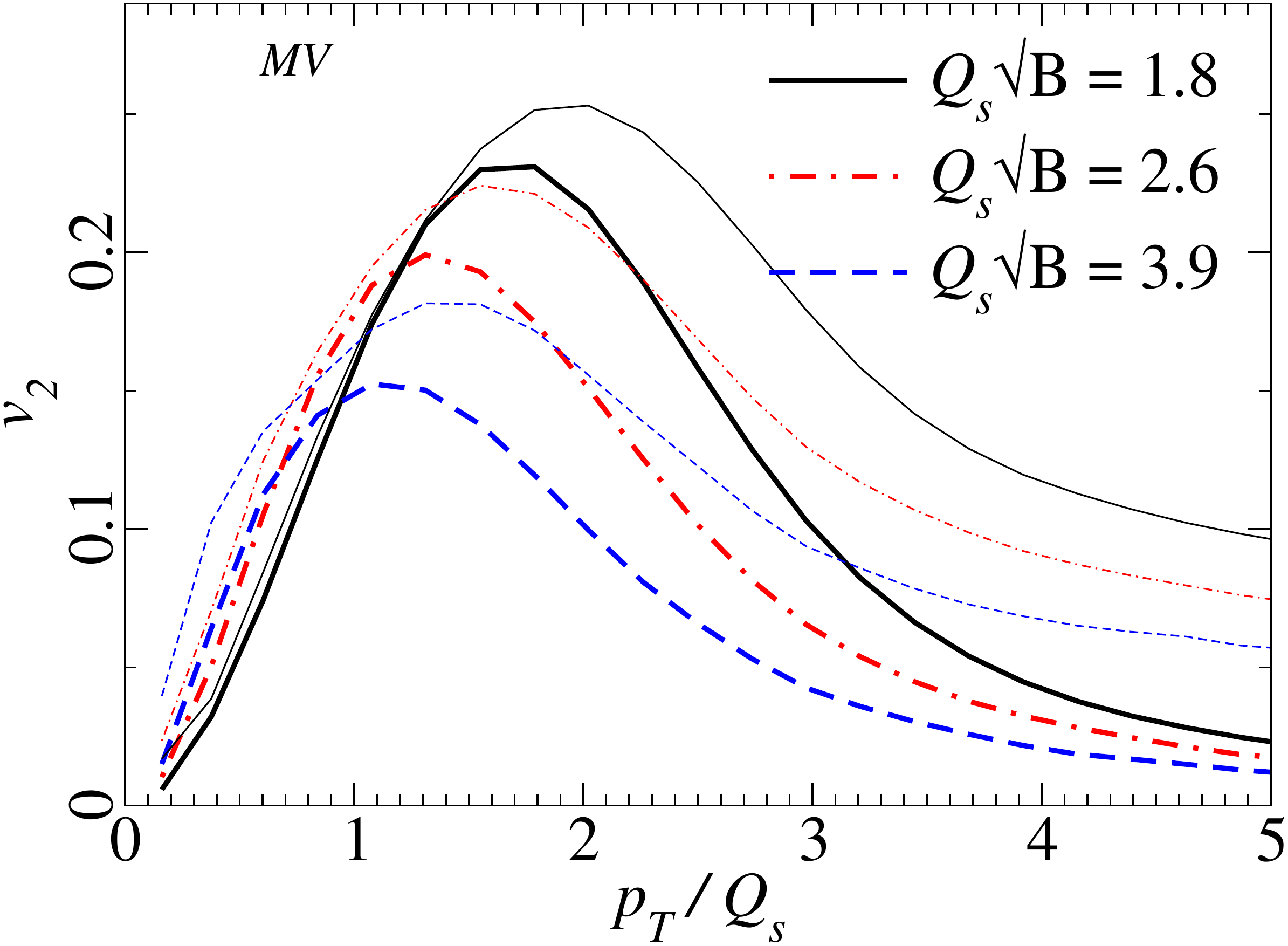}
\includegraphics[width=0.33\textwidth]{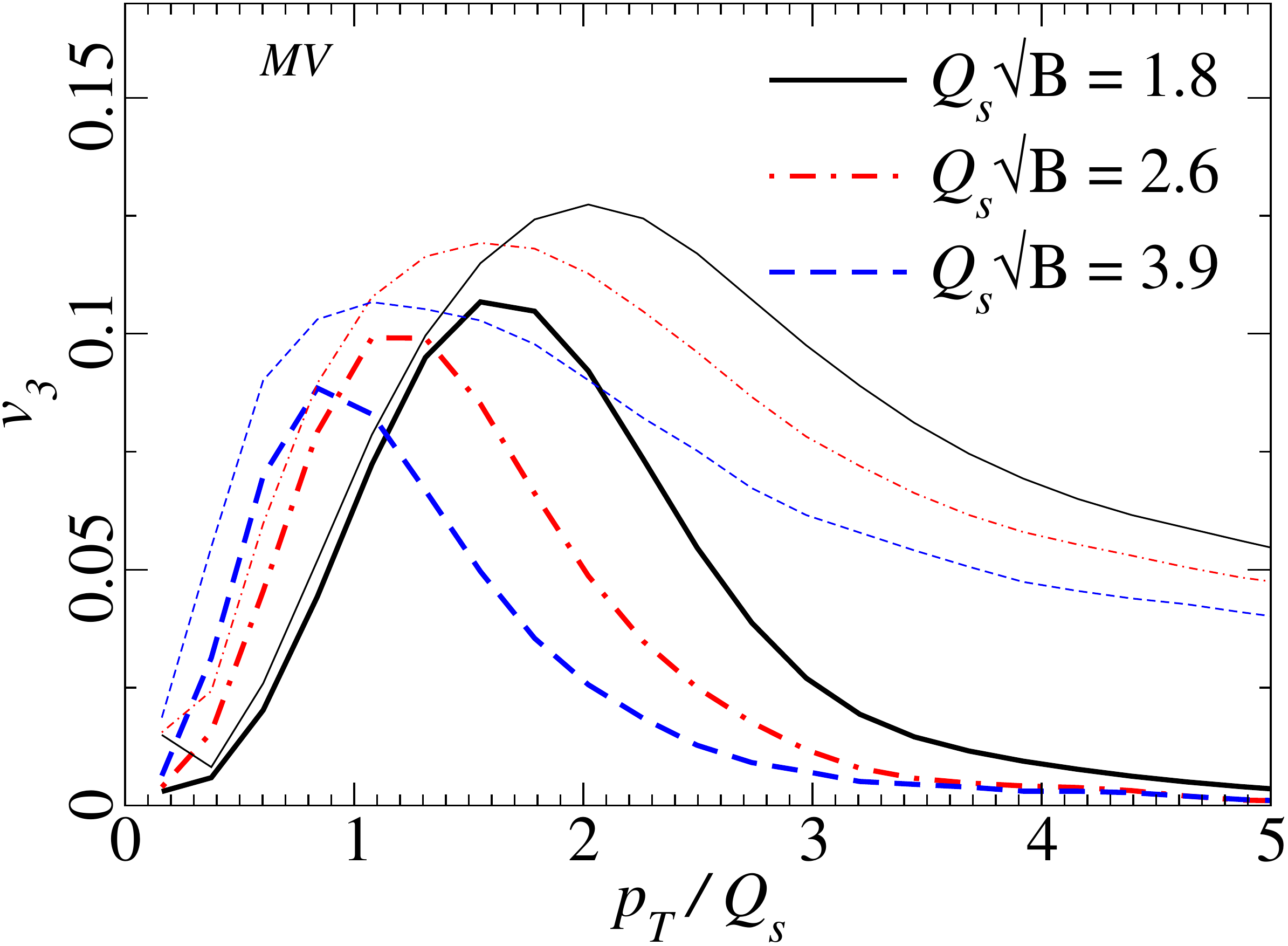}
\includegraphics[width=0.33\textwidth]{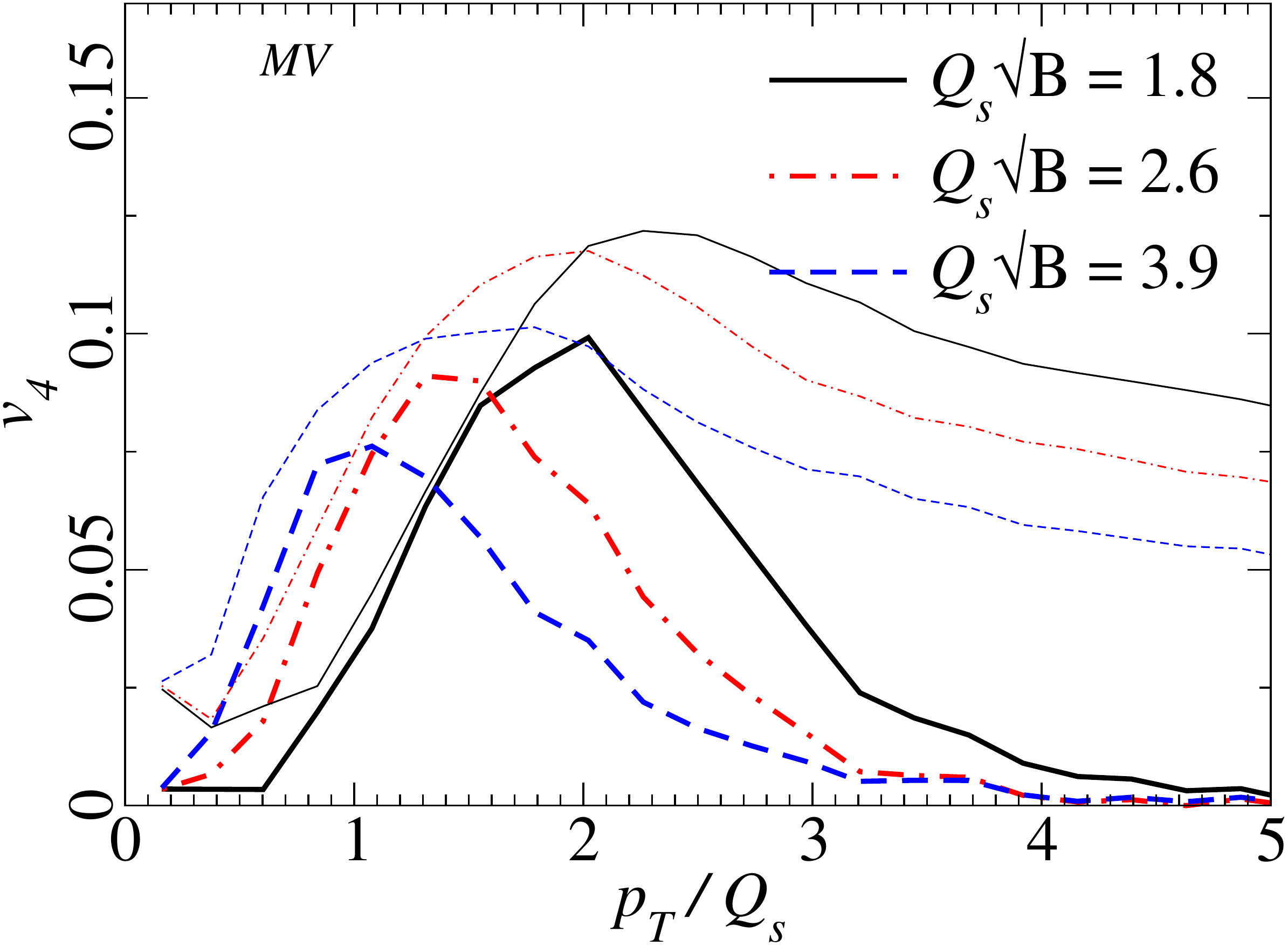}
}
\centerline{
\includegraphics[width=0.33\textwidth]{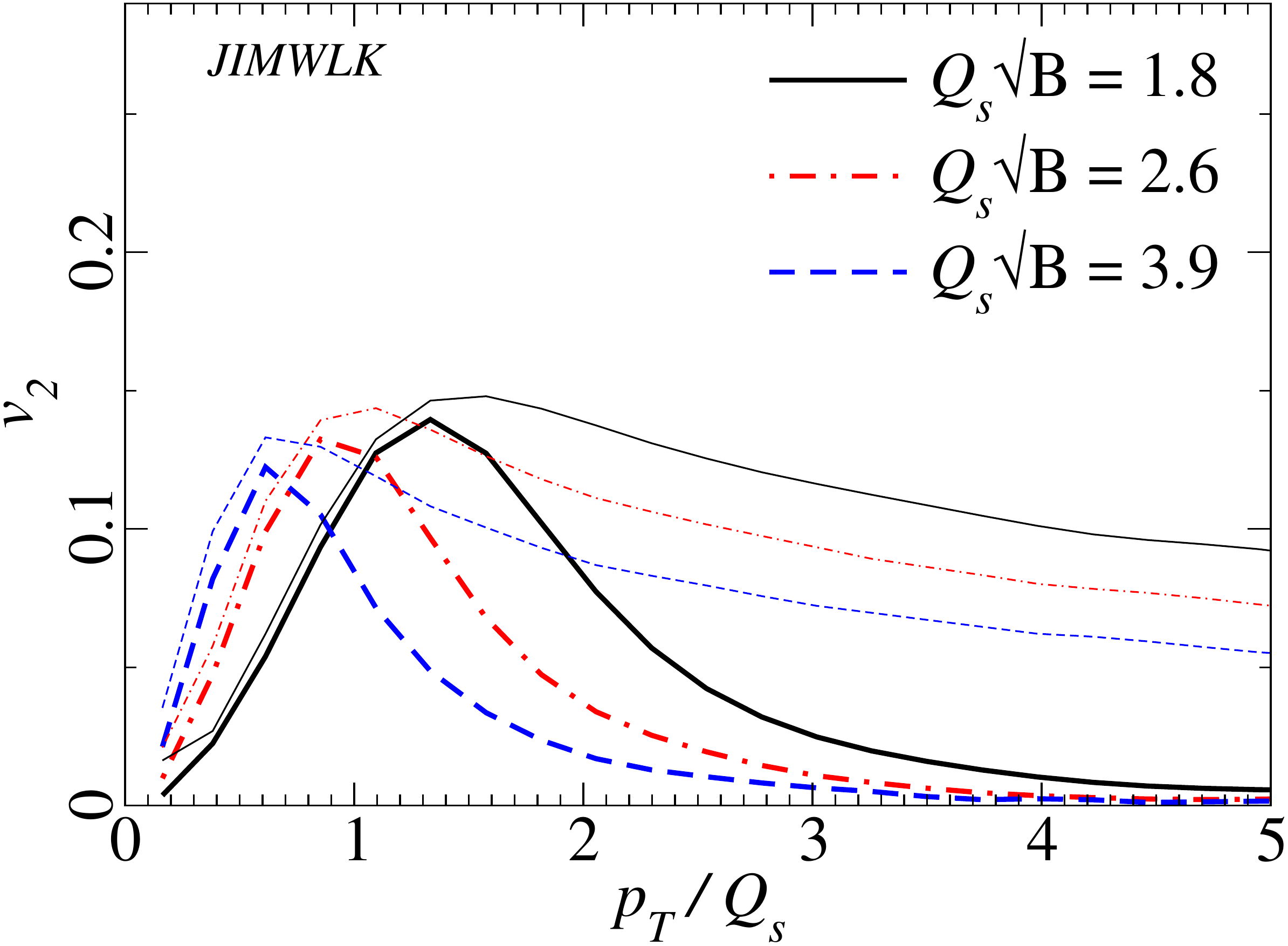}
\includegraphics[width=0.33\textwidth]{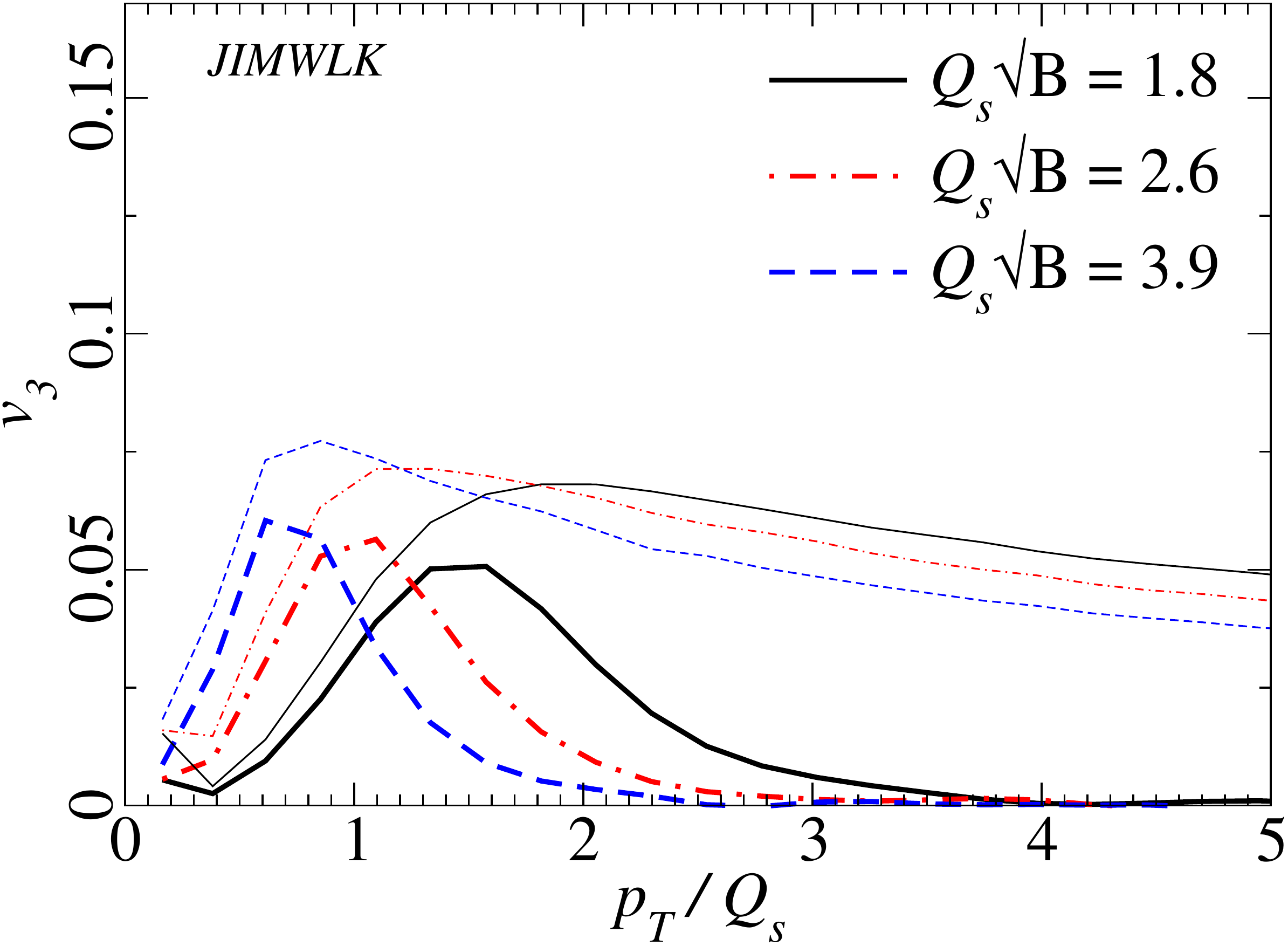}
\includegraphics[width=0.33\textwidth]{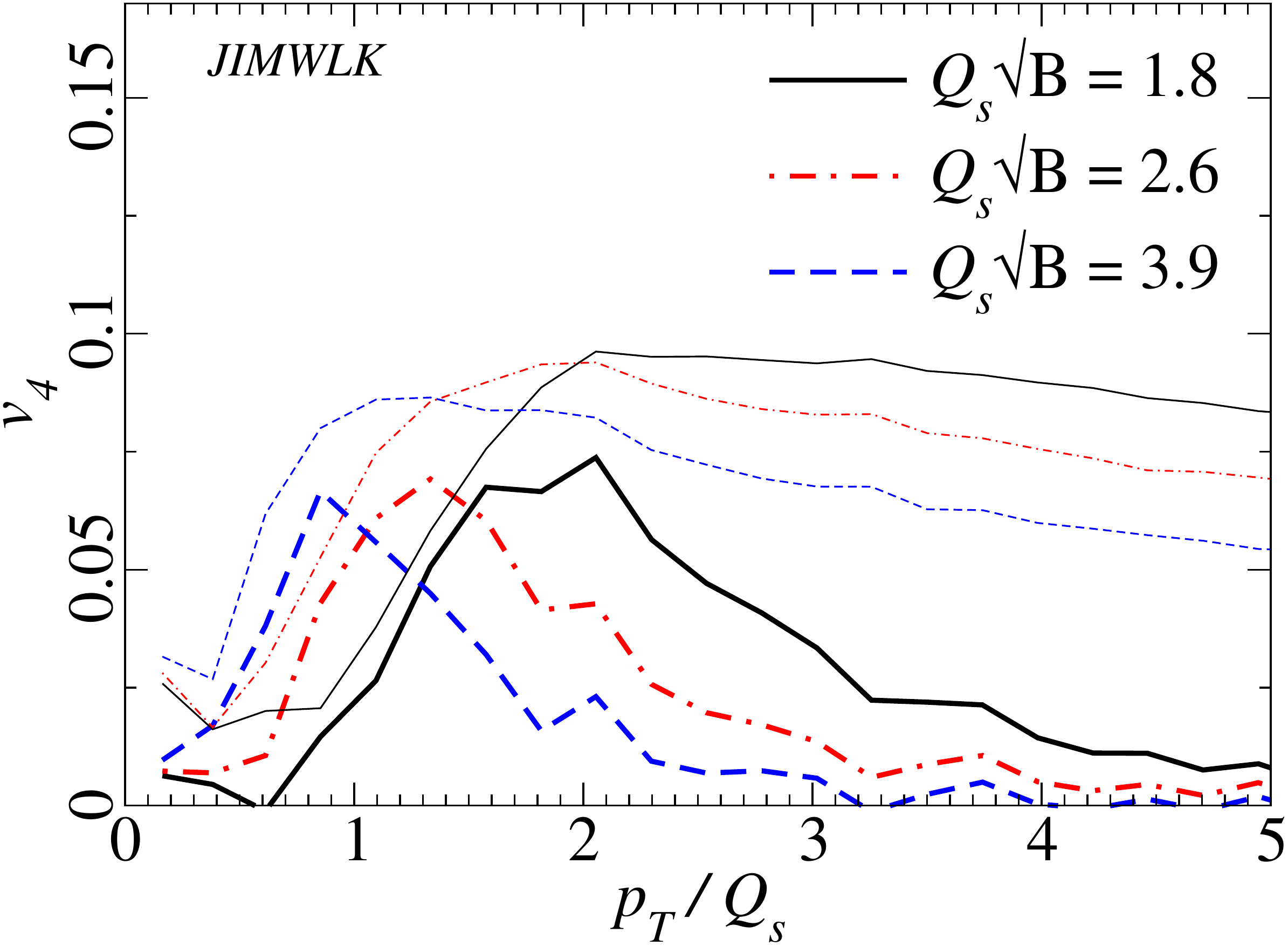}
}
\caption{Azimuthal harmonic coefficients $v_n\{2\}$ from \cite{Lappi:2015vha}, with, from left to right,  $n=2,3,4$. Top: Wilson lines from the MV dstribution. Bottom: Wilson lines from a numerical solution of the running coupling JIMWLK equation.  Thick lines:  particles correlated with reference particles at all transverse momenta. Thin lines: particles correlated only with reference particles in the same transverse momentum bin.}\label{fig:vns}
\end{figure}

The resulting anisotropy coefficients are shown in \fig\ref{fig:vns}. One sees that the transverse momentum dependence is very similar to the experimental data, rising first up to a transverse momentum a few times $\qs$ and then decreasing. The dependence on the size of the probe $B$ is clearly visible, with smaller probes exhibiting a stronger correlations. The values of $B$ and $\qs$ are chosen to be in the correct range for LHC energies. The results are shown both for two-particle correlations within the same transverse momentum bin and when a particle is correlated with reference particles at all momenta. There is a significant difference between the MV model and the JIMWLK-evolved distribution, pointing towards a surprisingly strong dependence of the result on the details transverse momentum distribution of gluons in the target. Similarly to the experimental data, the fourth harmonic $v_4$ peaks at a larger momentum than the second one. The calculation here is done with fundamental representation Wilson lines corresponding  to a quark probe, which leads to the existence of also a third harmonic $v_3$ component. For a gluon probe this  would be absent because the adjoint representation of SU(3) is real. We emphasize that the target here is completely isotropic and homogenous on average, and the effect is fully due to the fluctuations in the color field and not the geometrical shape of the target.

\section{Relating different approaches to target color field fluctuations}

Multiparticle correlations in the CGC picture have been studied in the literature using several different approaches.  We recently~\cite{Lappi:2015vta} explored more explicitly the relation between some of these approaches to the one of~\cite{Lappi:2015vha} described above.
The ``Glasma graph'' approximation has been used in a series of works~\cite{Dumitru:2010iy,Dusling:2012iga,Dusling:2013oia} to study the ridge correlation in proton-proton and proton-nucleus collisions.
 More recently Dumitru and collaborators have proposed an ``E-field domain model'' to explain the azimuthal asymmetries~\cite{Dumitru:2014dra}. 
Also the full dense-dense case has recently been studied~\cite{Schenke:2015aqa} with   numerical classical Yang-Mills evolution and a realistic target nucleus geometry.
It is important to realize that the physical picture of color field domains of size $1/\qs$ is the same in all of these calculations, and the differences are only in the approximations used. 

The relation between the calculations is easier to describe in terms of the color charge density $\rho$, which is related to the Wilson line by:
\begin{equation}
V(\xt)  = P \exp\left\{i g \int \ud x^- \frac{\rho(\xt,x^-)}{\nabt^2} \right\}.
\end{equation}\label{eq:rho}
The MV model assumes a Gaussian probability distribution for the color charges $\rho$.  It has been previously observed in the case of other correlations~\cite{Dumitru:2011vk} that also the probability distribution from JIMWLK evolution is very close to the Gaussian approximation. The recent calculation of Ref.~\cite{Lappi:2015vha} described above includes the full nonlinear dependence on the color charge $\rho$ resulting from the exponential \nr{eq:rho}. The glasma graph calculations differ from these in that the result has been linearized in $\rho$, which should be a good approximation for $\ptt \gtrsim \qs$. The CYM calculations~\cite{Schenke:2015aqa} also use the MV model and as such assume a Gaussian distribution in $\rho$.

The E-field domain model is formulated by expanding, as in the glasma graph approximation, the Wilson line correlator in the small dipole limit
\begin{equation}
 \frac{1}{\nc}V^\dag(\bt+\rt/2) V(\bt-\rt/2) 
\approx 1 - \frac{r^i r^j}{4\nc}E^a_i(\bt) E^a_j(\bt) .
\end{equation}
The difference with respect to the previous works is that the distribution of  color electric fields is assumed to have a very specific non-Gaussian form.
The two-point function of the electric field is taken to depend on a color field direction unit vector $\hat{a}$ as:
\begin{equation}
\left< E^j E^j\right> \sim \left[ \delta^{ij}(1-\mathcal{A}) + 2 \mathcal{A} \hat{a}^i\hat{a}^j \right].
\end{equation}
Expectation values of physical observables are then in the end obtained by averaging over the orientation of $\hat{a}$. This has the consequence that the four-point function of  electric fields $\left< E E E E \right>$ appearing in the two-particle correlation function has, in addition to the three contraction terms of a Gaussian distribution, an additional contribution that is proportional to $\mathcal{A}^2$. The magnitude of this additional parameter is not known, and one can present various interpretations for its physical origin.

In Ref.~\cite{Lappi:2015vta} we also performed a numerical comparison of the approximations discussed above (except for the E-field domain model which has the additional free parameter $\mathcal{A}$). We found that the ``nonlinear Gaussian'' approximation, taking the distribution of $\rho$'s to be Gaussian but treating the nonlinearity in \eq\nr{eq:rho} fully, is accurate to at least within 10\% for the azimuthal harmonics. Also the glasma graph approximation is in general very close to the full nonlinear JIMWLK result, but deviates from it by a factor of $~2$ for $v_2$ in the MV model.

\section{Conclusions}

In conclusion, there is by now ample theoretical evidence that fluctuations in the target color field do produce significant anisotropies directly in momentum space. These can be a significant contribution to the measured correlation signals for small systems, and should be taken into account together with effects of final state collectivity. Calculations of these correlations in different approximations in the CGC picture are rather consistent and are based on a common physical picture.  
For a more detailed comparison with experimental data the effects of hadronization should be added to these parton level calculations. In the future it would also be interesting to study in more detail the rapidity structure of the correlations~\cite{Iancu:2013uva}.

The author is supported by the Academy of Finland, projects
267321 and 273464. This research used computing resources of 
CSC -- IT Center for Science in Espoo, Finland.





\bibliographystyle{elsarticle-num}
\bibliography{spires}







\end{document}